\documentclass{iau} 

\usepackage{graphicx}
\usepackage{color}
\usepackage[colorlinks=true,citecolor=blue]{hyperref}
\usepackage[sort&compress,square,comma,authoryear]{natbib}

\newcommand{\oiii}{[O\,\textsc{iii}]}
\newcommand{\nii}{[N\,\textsc{ii}]}
\newcommand{\sii}{[S\,\textsc{ii}]}

\newcommand{\siii}{[S\,\textsc{iii}]}

\newcommand{\hb}{H$\beta$}
\newcommand{\ha}{H$\alpha$}

\title[The MAGNUM survey] %% give here short title %%
{Interstellar medium properties and feedback in local AGN with the MAGNUM survey}

\author[M. Mingozzi]   %% give here short author list %%
{M. Mingozzi$^1$,
 G. Cresci$^2$,
 G. Venturi$^{3,2}$,
 A. Marconi$^{4,2}$,
 F.~Mannucci$^2$}

\affiliation{$^1$INAF - Osservatorio astronomico di Padova, Vicolo dell'Osservatorio 5, 35122 Padova, Italy \\ email: {\tt matilde.mingozzi@inaf.it} \\[\affilskip]
$^2$INAF -- Osservatorio Astrofisico di Arcetri, Largo E. Fermi 5, I-50157, Firenze, Italy \\ $^3$ Instituto de Astrofísica, Pontificia Universidad Católica de Chile, Avda. Vicuña Mackenna 4860, 8970117, Macul, Santiago, Chile \\ $^4$ Dipartimento di Fisica e Astronomia, Università degli Studi di Firenze, Via G. Sansone 1, I-50019 Sesto Fiorentino, Firenze, Italy}

\pubyear{2020}
\volume{359}  %% insert here IAU Symposium No.
\jname{Galaxy evolution and feedback across different environments}
\editors{T. Storchi-Bergmann, R. Overzier, W. Forman \& R. Riffel, eds.}

\begin{document}

\maketitle

\begin{abstract}
We investigated the interstellar medium (ISM) properties in the central regions of nearby Seyfert galaxies characterised by prominent conical or bi-conical outflows belonging to the MAGNUM survey by exploiting the unprecedented sensitivity, spatial and spectral coverage of the integral field spectrograph MUSE at the Very Large Telescope. We developed a novel approach based on the gas and stars kinematics to disentangle high-velocity gas in the outflow from gas in the disc to spatially track the differences in their ISM properties. This allowed us to reveal the presence of an ionisation structure within the extended outflows that can be interpreted with different photoionisation and shock conditions, and to trace tentative evidence of outflow-induced star formation (``positive'' feedback) in a galaxy of the sample, Centaurus~A.
%% add here a maximum of 10 keywords, to be taken form the file <Keywords.txt>
\end{abstract}

\firstsection % if your document starts with a section,
              % remove some space above using this command.
\section{Introduction} 
Galaxy-scale outflows driven by active galactic nucleus (AGN) activity are thought to be so powerful to sweep away most of the gas of the host galaxy, providing a mechanism for the central black hole (BH) to possibly regulate star formation (SF) activity. This mechanism, the so-called negative feedback, could potentially explain the relation between the BH mass and the galaxy bulge properties \citep{silk1998,fabian2012}.
Recently, models and observations have revealed that outflows and jets can also have a positive feedback effect, triggering SF in the galaxy disc and also within the outflowing gas itself (e.g. \citealt{silk2013,cresci2015b,maiolino2017}).
Outflows are now routinely detected in luminous active galaxies on different physical scales and in different gas phases (e.g., ionised, atomic and molecular gas; \citealt{cicone2018} and references therein), even though understanding their role in galaxy evolution is still a challenging task. 
In this context, nearby galaxies represent ideal laboratories to explore in high detail outflow properties, their formation and acceleration mechanisms, as well as the effects of SF and AGN activities on host galaxies.

Here we present the results of the Measuring AGN under MUSE microscope (MAGNUM) survey (P.I. Marconi), aimed at investigating the inner regions of a number of local AGN, all showing evidence for the presence of outflows, with the unprecedented combination of spatial and spectral coverage of the integral field spectrograph MUSE \citep{bacon2010} at the Very Large Telescope. 
This contribution is based on recent published papers \citep{cresci2015b,venturi2017,venturi2018,mingozzi2019} and on unpublished material from M. Mingozzi's Phd Thesis (2020, University of Bologna; \href{http://amsdottorato.unibo.it/9461/1/thesis_final_mingozzi.pdf}{link}).

\section{The MAGNUM survey}
MAGNUM galaxies have been selected to be observable from Paranal Observatory and with a luminosity distance $D_{L} < 50$~Mpc. In \textcolor{blue}{Venturi et al. (2020, in preparation)} we present our sample, explaining the selection criteria, data reduction and analysis, and investigating the kinematics of the ionised gas. Here, we show our results for the nine Seyfert galaxies analysed in \citet{mingozzi2019} (M19 hereafter), namely Centaurus~A, Circinus, NGC~4945, NGC~1068, NGC~1365, NGC~1386, NGC~2992, NGC~4945 and NGC~5643. The MUSE field of view (FOV) covers their central regions, spanning from 1 to 10 kpc, according to their distance. The average seeing of the observations is $\sim0.6$"$-0.8$". 
%The data reduction was performed using the MUSE pipeline (v1.6). %The final datacubes consist of $\sim 300 \times 300$ spaxels, for a total of over 90000 spectra with a spatial sampling of 0.2"~$\times$~0.2" and a spectral resolution going from 1750 at 4650 \AA \,to 3750 at 9300 \AA. 
The datacubes were analysed with a set of custom python scripts in order to fit and subtract the stellar continuum in each single-spaxel spectrum and fit the main emission lines (i.e. H$\beta$, \oiii$\lambda\lambda$4959,5007, H$\alpha$, \nii$\lambda\lambda$6548,84, \sii$\lambda\lambda$6717,31, \siii$\lambda$9069) with multiple Gaussians where needed. This happens in the central parts of the galaxies and in the outflowing cones. All the details about the applied procedure are given in M19.% and in \textcolor{blue}{Venturi et al. (2020, in preparation)}. 

\subsection{Gas properties: disc versus outflow}
\begin{figure}
\centering
% \vspace*{-2.0 cm}
\begin{center}
 \includegraphics[width=2.in]{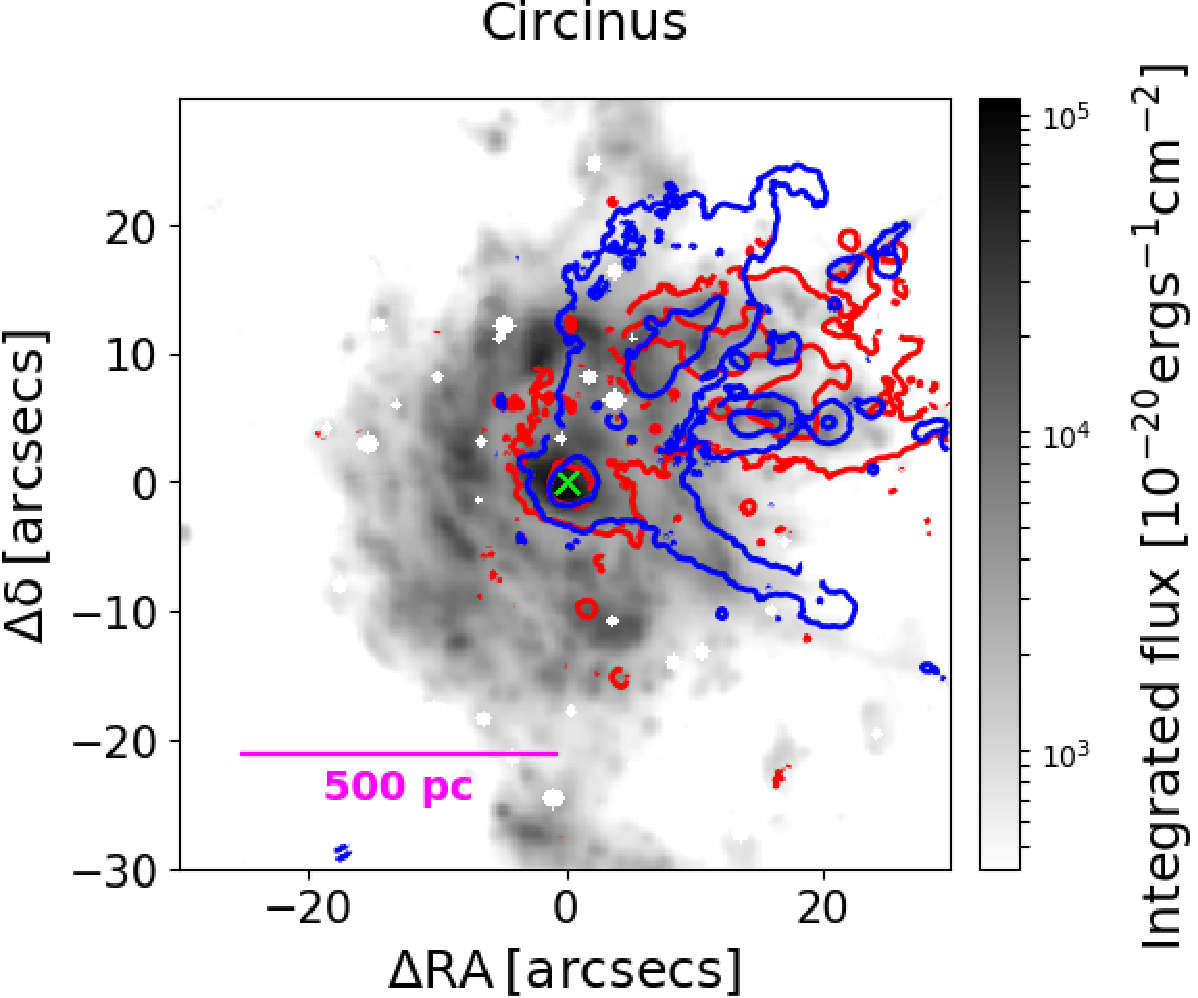} 
 \vspace{0.1cm}
 \includegraphics[width=2.in]{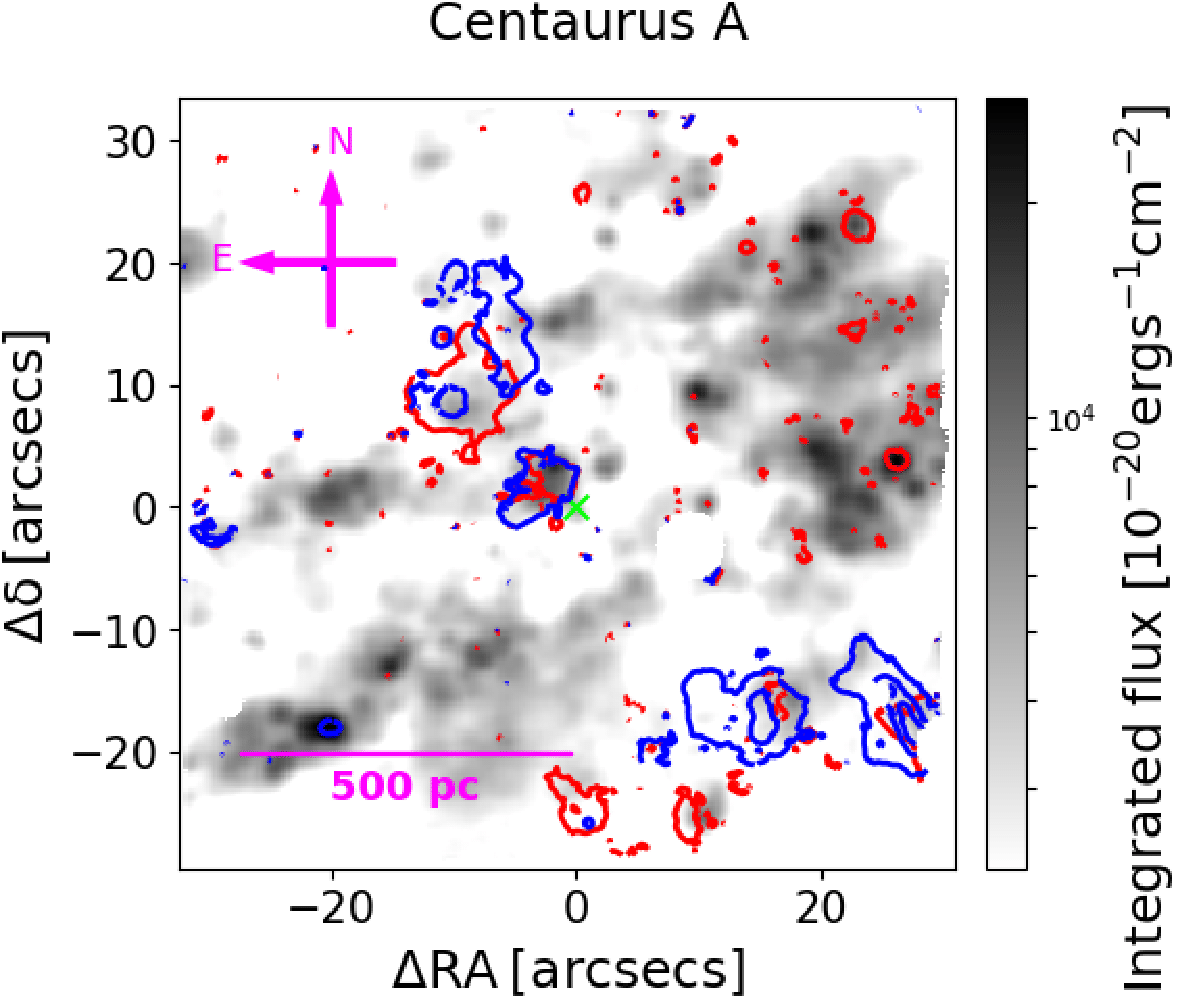} 
% \vspace*{-1.0 cm}
 \caption{Circinus and Centaurus~A H$\alpha$ disc component maps with \oiii$\lambda5007$ blueshifted and redshifted outflow component contours superimposed (in blue and red, respectively). We show only the spaxels with a signal-to-noise S/N~$>5$. East is to the left. The magenta bar represents a physical scale of $\sim500$~pc. The white circular regions are masked foreground stars. The green cross marks the position of the peak of the continuum in the wavelength range $6800-7000$~\AA.}
  \label{fig:vbin}
\end{center}
\end{figure}

% Taking advantage of integral field spectroscopy, excitation, ionisation conditions and dust attenuation of the ISM across the MUSE FOV of MAGNUM galaxies can be investigated through specific emission-line ratios, used as diagnostics to infer the properties of the disc and the outflow.
In many works, disc and outflow are separated according to the width of the two Gaussian components used to fit the main emission lines (narrower and broader, respectively). In our analysis this approach is not feasible since the line profiles can be very complex, requiring three or four Gaussians to be fully reproduced. Therefore, we disentangle the outflow from the systemic gas by applying a novel approach, explained in detail in M19. 
In brief, we assume that the stellar velocity is generally a good approximation of the gas velocity in the disc, defining as \textit{disc component} the low-velocity ionised gas rotating similarly to the stars, while the \textit{outflow component} (i.e. the high-velocity component) is moving faster than the stellar velocity, and is partly blueshifted and partly redshifted with respect to it.
% To do this, we define velocity channels of $\sim50$~km/s, ranging from $-1000$~km/s to $+1000$~km/s around the fitted emission lines, centring the zero velocity to the measured stellar velocity in each spaxel of the MUSE FOV for all the galaxies and associating the velocity channels close to the core of the lines in the fitted line profile with the disc (\textit{disc component}), and the sum of the redshifted and blueshifted channels with the outflow (\textit{outflow component}). The flux in each bin is obtained by integrating the fitted line profile within each velocity channel of the emission lines taken into account.
% Since Centaurus~A is characterised by a strong misalignment between stars and gas (e.g. \citealt{morganti2010}), we consider the global systemic velocity ($v_{\rm{sys}}=547$~km/s) as a reference for the gas disc velocity in this object.
As an example, Fig.~\ref{fig:vbin} shows the H$\alpha$ disc component flux maps, superimposing the \oiii$\lambda5007$ outflow component contours for Circinus and Centaurus~A\footnote{Centaurus~A shows a strong misalignment between stars and gas (\citealt{morganti2010}), so we consider the global systemic velocity ($v_{\rm{sys}}=547$~km/s) as a reference for the disc gas.}. Indeed, the H$\alpha$ emission is in general dominant in the disc, while the \oiii\ is enhanced in the outflow \citep{venturi2018}.
In Circinus (1"~$\sim 20.4$~pc), the outflow is extended on North-West in one-sided and wide-angled kpc-scale \oiii \, cone, first revealed by \citet{marconi1994}. In Centaurus~A (1"~$\sim 18.5$~pc) the outflow is mainly distributed in two cones (direction north-east and south-west) in the same direction of the extended double-sided jet revealed both in the radio and X-rays (e.g. \citealt{hardcastle2003}), and located perpendicularly with respect to the gas in the disc component.

\begin{figure}[b]
\centering
 \includegraphics[width=1.5in]{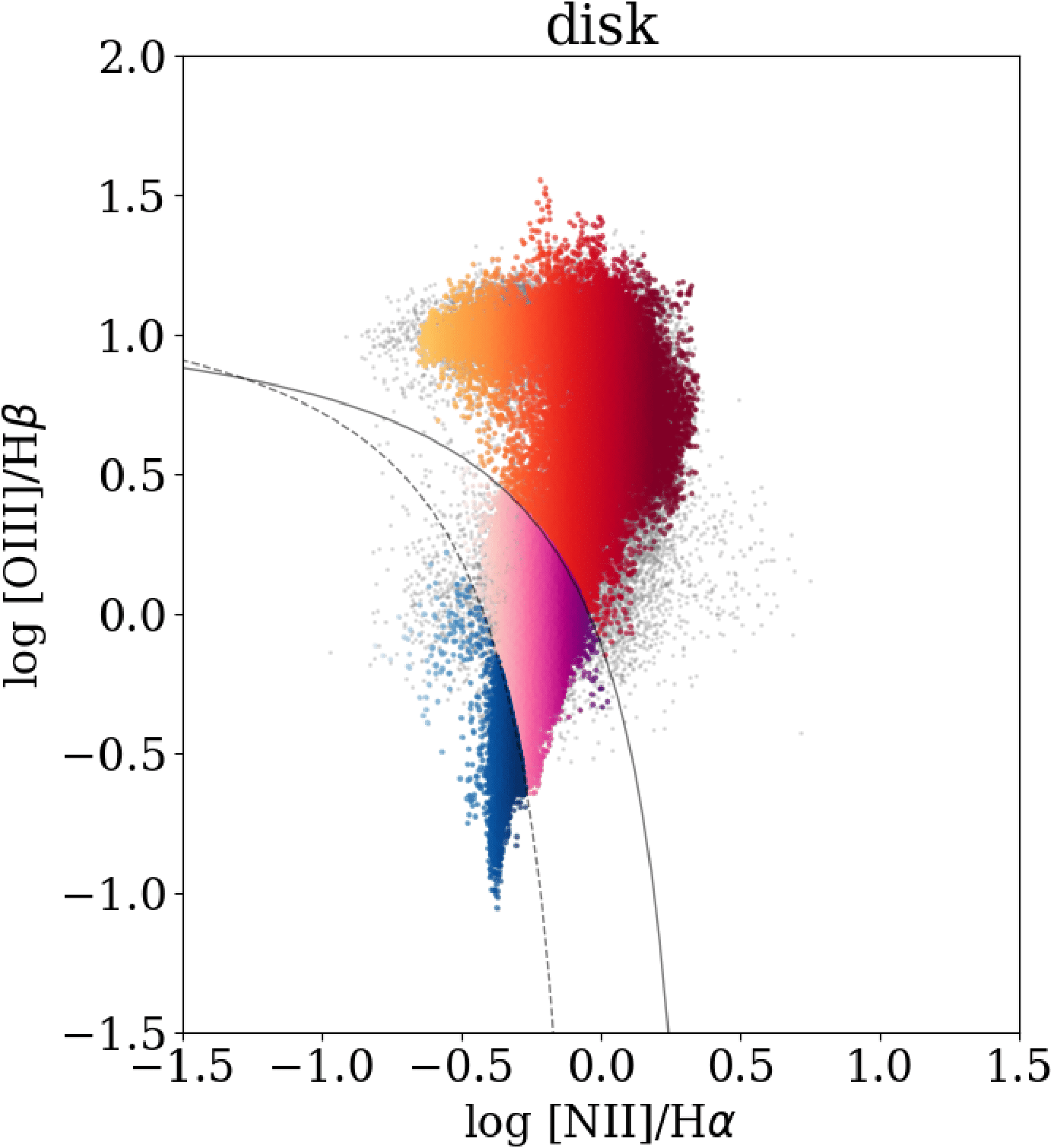} 
 \includegraphics[width=1.5in]{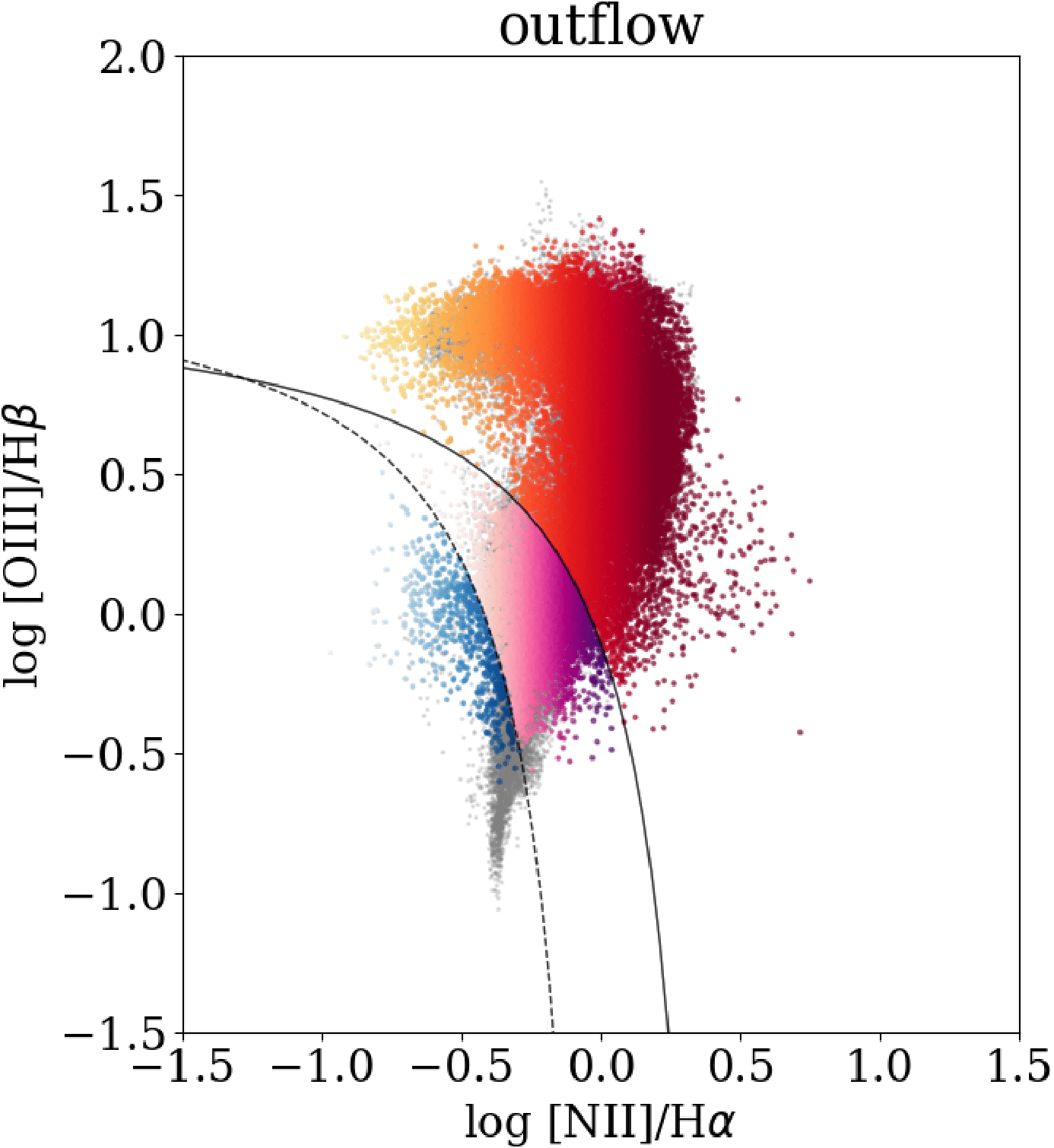} 
 \includegraphics[width=1.6in]{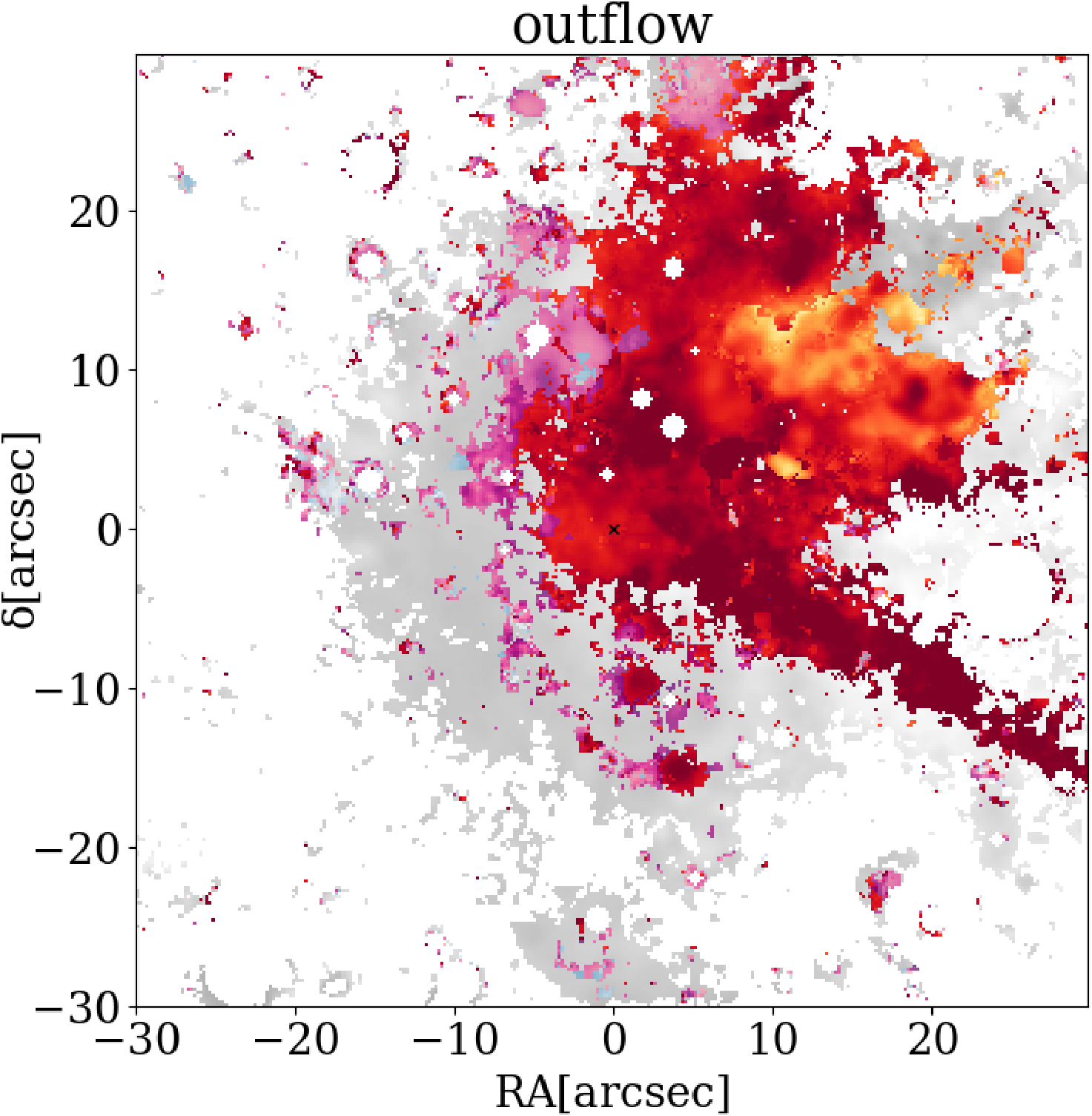}
 \caption{\nii-BPT diagrams for the disc and outflow components of Circinus, and the corresponding outflowing gas component map. Shades of blue, pink and red denote SF, composite and AGN dominated regions, respectively (darker shade means higher \nii/\ha). The black dashed curve is the boundary between star-forming galaxies and AGN \citep{kauffmann2003}, while the black solid curve is the theoretical upper limit on SF line ratios \citep{kewley2001}. The grey dots in the BPTs and the dashed grey regions in the corresponding map show the disc and outflow component together. We show only spaxels with a S/N~$>5$ for all the lines involved.}
\label{fig:cir_bpt}
\end{figure}

In M19, we calculated dust extinction, gas density and ionisation parameter (i.e. a measure of the radiation field intensity, relative to gas density) for the disc and outflow components, using \ha/\hb, \sii$\lambda$6717/\sii$\lambda$6731 and \siii$\lambda\lambda$9069,9532/\sii$\lambda\lambda$6717,31 line ratios, respectively.
We found that the outflow is characterised by higher values of density and ionisation parameter ($ A_V \sim0.9$, $ n_e \sim 250$~cm$^{-3}$, log(\siii/\sii)~$\sim 0.16$) than the disc component, that is instead more affected by dust extinction ($ A_V \sim1.75$, $ n_e \sim 130$~cm$^{-3}$, log(\siii/\sii)~$\sim -0.38$). Interestingly, our median outflow density is lower than what is found in literature, but a more consistent value can be obtained calculating the median density weighting by the \sii\, line flux (disc: $n_e \sim 170$~cm$^{-3}$; outflow: $ n_e \sim 815$~cm$^{-3}$). This means that many values of outflow density found in literature could be biased towards higher densities because they are based only on the most luminous and densest outflowing regions, characterised by a high S/N ratio. 
Finally, in M19 we used spatially and kinematically resolved Baldwin-Phillips-Terlevich (BPT, \citealt{baldwin}) diagrams to explore the dominant contribution to ionisation in the disc and outflow components separately. 
The left and middle panels of Fig.~\ref{fig:cir_bpt} show the \nii-BPT diagrams of Circinus for the disc and outflow, respectively. %: shades of blue, pink and red denote SF, composite and AGN dominated regions (darker shades means higher \nii/\ha), respectively. 
%The LI(N)ER-like excitation can be due either to shock excitation (e.g. \citealt{dopita1995}) or to hard-X radiation coming from AGN and to hot evolved (post-asymptotic giant branch) stars (e.g. \citealt{belfiore2016}). 
The corresponding position on the outflowing gas component map is shown in the right panel of Fig.~\ref{fig:cir_bpt}. %In the background of all the pictures (black dots in the BPTs and shaded grey in the corresponding maps), we show the disc and outflow components together to allow us a better visual comparison.
We noticed that %the outflow spans a wider range of \nii/H$\alpha$, including lower and higher values compared to the disc. Similar results are obtained also for the \sii-BPT. Specifically, 
the highest and lowest values of low-ionisation line ratios (LILrs; i.e. \nii/H$\alpha$ and \sii/H$\alpha$), displayed in dark red and orange, are prominent in the AGN/LI(N)ER-dominated outflow component, while they are not observable in the disc component. 
These features are visible in almost all the MAGNUM sample: in general, the highest LILrs trace the inner parts along the axis of the emitting cones, where the \siii/\sii\, line ratio is enhanced (i.e. high ionisation), while the lowest LILrs follow the cone edges and/or the regions perpendicular to the axis of the outflow, characterised also by a higher \oiii\, velocity dispersion (M19). 
A possible explanation for the observed features identified in the outflow is to take into account different proportions of two distinct populations of line emitting clouds (e.g. \citealt{binette1996}). One is optically thin to the radiation and characterised by the highest excitation, while the other, optically thick, is impinged by a filtered - and then harder - radiation field, which makes it dominated by low-excitation lines and characterised by lower \siii/\sii\, line ratios. The highest LILrs may be due to shocks and/or to a hard-filtered radiation field from the AGN (M19).

\subsection{Centaurus A: a local laboratory to study AGN positive feedback}
\begin{figure}[b]
\begin{center}
 \includegraphics[width=5in]{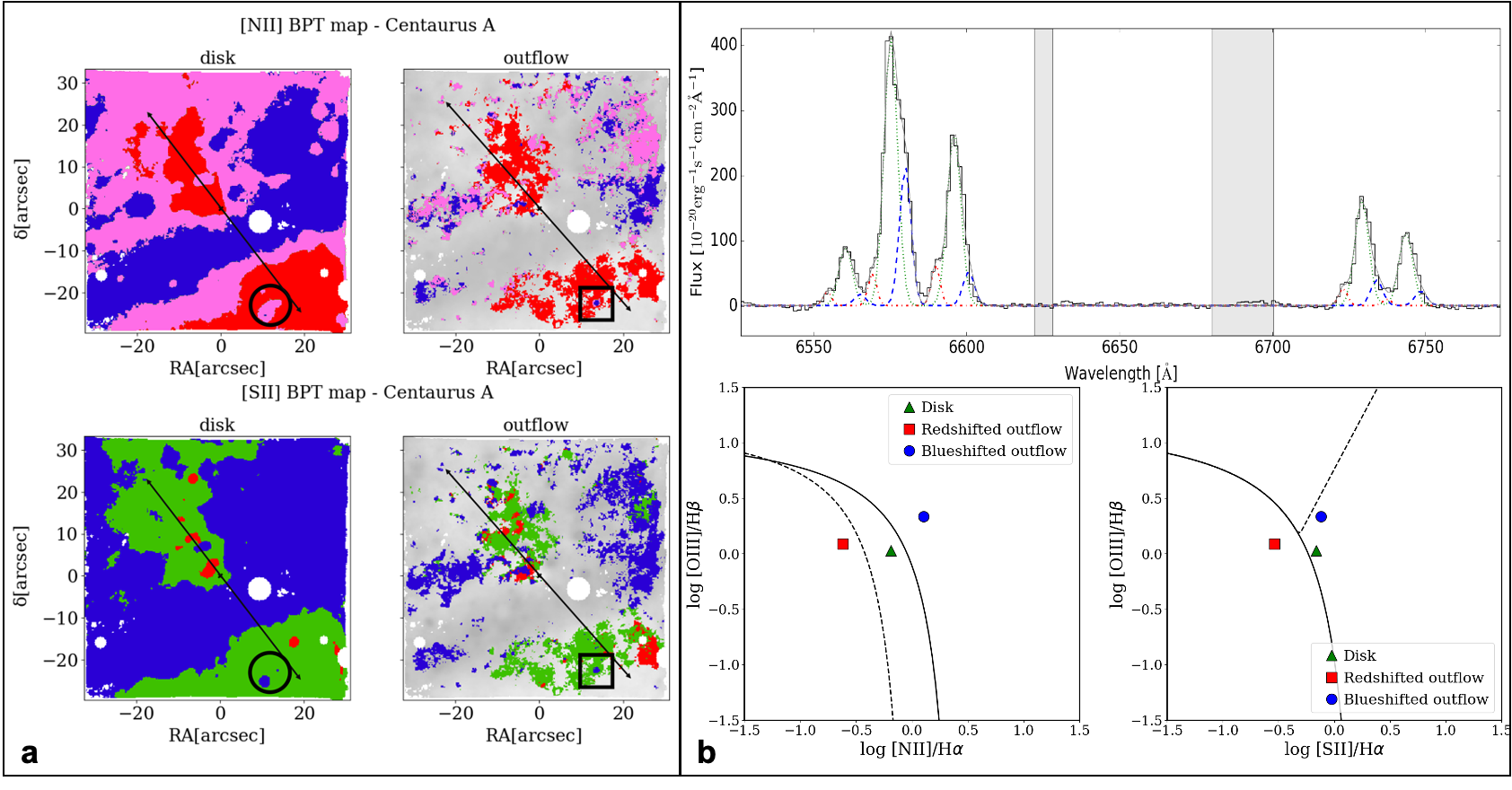} 
 \caption{\textbf{a}) \nii\, (top panel) and \sii\, (bottom panel) BPT maps for Centaurus~A disc (-150~km/s $<$ v $< $ + 150~km/s) and outflow components (v $>$ + 200 km/s and v $<$ - 200 km/s). SF, AGN, Composite and LI(N)ER dominated regions are indicated in blue, red, pink and green, respectively. The black arrow indicates the jet direction (see Fig. 6 in \citealt{hardcastle2003}), while the solid circle and square highlight the position of composite/SF blobs, located in the outflow direction. \textbf{b}) Spectrum of the SF blob from the outflow flux maps in the wavelength range $6500-6650$~\AA (top panel), fitted by a three Gaussian component fit (dashed-dotted blue, dotted green and dashed red Gaussians for the blueshifted, systemic and redshifted components, respectively) and their corresponding \nii- and \sii-BPT diagrams (bottom panel).}
  \label{fig:bpt_cena}
\end{center}
\end{figure}

Centaurus~A shows %a characteristic double-sided jet observed in radio and X-rays (e.g. \citealt{israel1998,morganti2010}) and is 
the best example of a radio jet emitted by the central AGN interacting with the ISM (e.g. \citealt{santoro2016} and references therein). 
In this context, we investigated its central region, exploiting our approach of disentangling the disc from the outflow component to obtain an independent classification of their ionisation sources. %with the BPT diagrams. 
Fig.~\ref{fig:bpt_cena}\textbf{a} shows the \nii- and \sii-BPT maps for the disc (on the left) and the outflow (on the right). %, with regions dominated by SF, AGN, Composite and LI(N)ER activity indicated in blue, red, pink and green, respectively. 
%The direction of the inner jet (see Fig. 6 in \citealt{hardcastle2003}), consistent with the location of the bi-conical outflow dominated by LI(N)ER- and AGN-like ionisation, is marked by the black arrow. %An \oiii\, ionisation cone of $\sim500$~pc along the jet axis, related to AGN or shock activities, was revealed for the first time by \citet{bland2003}. Moreover, \citet{israel2017} discovered an outflow of gas that from the center of CenA extends along an axis close to that of the northern X-ray/radio jet, detected in the neutral gas traced by CO and and \ci, and in the far-infrared through fine-structure lines. \citet{israel2017} detected also a less clear southern counterpart to the well-collimated northern outflow. 
The galaxy FOV South-West portion hosts a blob (solid circle) with a velocity consistent with the gas disc and by Composite and SF dominated ionisation (in \nii- and \sii-BPT, respectively), that can be interpreted as SF triggered in the ionisation cone due to compression of the galaxy ISM by the outflow, as already revealed in another galaxy of the MAGNUM survey, NGC~5643, by \citet{cresci2015b}. Moreover, a nearby clump (solid square) appears to have SF ionisation, but velocities consistent with the gas in the outflow, suggesting that newborn stars could be forming directly in outflowing gas (see \citealt{maiolino2017}). %If confirmed, this would be the first example of these two types of positive feedback coexisting in the same object.
% \begin{figure}
% % \vspace*{-2.0 cm}
% \begin{center}
%  \includegraphics[width=3.4in]{Picture_1.png} 
%  \vspace{0.1cm}
%  \includegraphics[width=3.4in]{Picture_2.png} 
% % \vspace*{-1.0 cm}
%  \caption{\nii\, (upper panel) and \sii\, (lower panel) BPT maps for the disc (-150~km/s $<$ v $< $ + 150~km/s) and outflow component (v $>$ + 200 km/s -- v $<$ - 200 km/s) of CenA. SF, AGN, Composite and LI(N)ER dominated regions are indicated in blue, red, pink and green, respectively. The black arrow indicates the direction of the outflow, while the solid circle and square highlight the position of composite/SF blobs, located in the direction of the outflow.}
%   \label{fig:bpt_cena}
% \end{center}
% \end{figure}
The top panel of Fig.~\ref{fig:bpt_cena}\textbf{b} shows the spectrum of the star-forming blob in the outflow (solid square in Fig.~\ref{fig:bpt_cena}\textbf{a}). %, clearly presenting outflow signatures. 
The asymmetric line profiles can be reproduced by a fit with three Gaussian components, whose corresponding positions in \nii- and \sii-BPT diagrams are shown in the bottom panel of Fig.~\ref{fig:bpt_cena}\textbf{b}: the redshifted Gaussian component (red) has a very strong \ha\ emission (star formation rate SFR~$\sim 8\times10^{-3}$~M$_{\odot}$/yr, using \citealt{lee2009} calibration) and SF ionisation.
Overall, the two blobs have a total SFR~$\sim0.01$~M$_{\odot}$/yr, which is $\sim 3$\% of the global value of the galaxy % (SFR~$=0.4$~M$_{\odot}$/yr \citealt{diamond-stanic2012}), 
and could represent the first evidence of both the two modes of positive feedback -- triggered SF both in the galaxy disc and within the outflowing gas -- operating in the nuclear region of this galaxy.
%However, \citet{maiolino2017} demonstrated that only the combination of X-Shooter \citep{dodorico2006,vernet2011} and MUSE data allow to undoubtedly identify the signatures of SF in the outflow. Indeed, X-shooter has a broad-band spectral coverage able to cover IR diagnostics, such as the \feii$\lambda1.64\mu$m/Br$\gamma\lambda2.16\mu$m versus the H$_2(1-0)$S(1)$\lambda2.12\mu$m/Br$\gamma$ and the \feii$\lambda1.25\mu$m/Pa$\beta\lambda1.28\mu$m versus \pii$\lambda1.18\mu$m/Pa$\beta$ diagrams \citep{oliva2001,colina2015}, that can further discriminate among SF, AGN and shock ionisation. % Typically, in star forming regions these ratios are expected to be very low, since shocks could enhance the emission of \feii\, and H$_2$, while the Br$\gamma$ and the \pii\, are direct tracers of the AGN UV and soft-X ionising radiation, respectively.
%Moreover, X-shooter high spectral resolution allows to investigate the presence of young stars formed in the outflow. To do this, the stellar absorption features, such as the Balmer series down to $\lambda \sim 4000$~\AA \,(tracing young hot O-B-type stars), the Ca~II triplet at $\lambda \sim 8500$~\AA \,(in case of recent SF, dominated by young red supergiants and young asymptotic giant branch stars) and, possibly, the weak absorption feature of HeI$\lambda4922$~\AA (unambiguous tracer of B-type stars and of young stellar populations) can be taken into account.
Hence, we asked for X-SHOOTER observations (0102.B-0292, P.I. Mingozzi), that we will discuss in a forthcoming paper \textcolor{blue}{(Mingozzi et al. 2020, in preparation)} to undoubtedly identify SF signatures in the outflow, exploiting IR diagnostics to discard AGN and shock ionisation, and stellar absorption lines to investigate the presence of newborn stars (see \citealt{maiolino2017}).

% \section{Conclusions}
% We explored the gas properties (e.g. density, ionisation parameter, reddening and source of ionisation) of the outflowing gas in the (E)NLR of the 9 nearby Seyfert galaxies that are part of the MAGNUM survey, all characterised by prominent conical or biconical outflows. Exploiting the very high spatial resolution of the optical integral field MUSE spectrograph at VLT, we were able to disentangle the outflow component from the disc component in order to analyse its peculiarities through spatially and kinematically resolved maps. To do this, we divided the main emission lines (H$\beta$, \oiii, \oi, H$\alpha$, \nii, \sii\, and \siii) in velocity bins, associating the core of the lines (centred on the stellar velocity in each spaxel) with the disc, and the blueshifted and redshifted wings with the outflow.

\section{Conclusions}\label{sec:magnum_conclusions}
The MAGNUM survey is exploring gas properties and ionisation sources of the outflowing gas in the central regions of nearby Seyfert galaxies.
We found that the gas in the outflowing cones of our galaxies is set up in clumpy clouds characterised by higher density and ionisation with respect to disc gas. The cone innermost regions are generally highly ionised and directly heated by the AGN. The cone edges and the regions perpendicular to the outflow axis could instead be dominated by shocks due to the interaction between the outflow and the ISM. Alternatively, these regions, generally characterised by low ionisation, could be impinged by an ionising radiation filtered by clumpy, ionised absorbers.
Separating the outflow and disc components allowed us also to detect in one of the sources, Centaurus~A, two blobs dominated by SF ionisation apparently embedded in the AGN ionisation cone, possibly tracing positive feedback (both in the disc and in the outflow) and accounting for $\sim 3$\% of the galaxy global SFR. If the new X-SHOOTER data confirmed it, this would be the first example of the two modes of positive feedback coexisting in the same object. The contribution to the total SFR might seem irrelevant, but it remains that Centaurus~A could be a local test bench to explore in detail this phenomenon. Positive feedback may play a significant role in the formation of galaxy spheroidal component at high redshift, where AGN-driven outflows are more prominent and the associated SF inside those very massive outflows possibly far higher \citep{gallagher2019, delpino2019}.
%the kinematics of the gas, improving our knowledge on positive feedback and possibly confirming that it is not a rare mechanism. This phenomenon may play a significant role in the formation of the spheroidal component of galaxies at high redshift, where AGN-driven outflows are more prominent and the associated SF inside those very massive outflows much higher \citep{gallagher2019, delpino2019}.

% \bibliography{references} 

\end{document}